\documentclass{aa}

\usepackage{graphicx}
\usepackage{txfonts}
\newcommand{\dd}{\mathrm{d}}

\usepackage{xcolor}

\usepackage{ulem}

\usepackage{aas_macros}
\usepackage{url}
\usepackage{hyperref}
\usepackage{natbib}
\def\gsim{\lower.5ex\hbox{$\; \buildrel > \over \sim \;$}}
\def\lsim{\lower.5ex\hbox{$\; \buildrel < \over \sim \;$}}
\begin{document}

   \title{Diffusive shock acceleration of dust grains at supernova remnants}


   \author{P. Cristofari
          \inst{1},
          V. Tatischeff\inst{2}, 
         and M. Chabot\inst{2} 
          }

   \institute{Laboratoire Univers et Théories, Observatoire de Paris, Université PSL, Université Paris Cité, CNRS, F-92190 Meudon, France\\
              \email{pierre.cristofari@obspm.fr}
         \and
            Universit\'{e} Paris-Saclay, CNRS/IN2P3, IJCLab, 91405 Orsay, France\\
             }

   \date{Received September 15, 1996; accepted March 16, 1997}

 
  \abstract
   {Diffusive shock acceleration (DSA) is a prominent mechanism for energizing charged particles up to very large rigidities at astrophysical collisionless shocks. In addition to ions and electrons, it has been proposed that interstellar dust grains could also be accelerated through diffusive shock acceleration, for instance, at supernova remnants (SNRs).}
   { Considering interstellar dust grains of various size and composition, we investigate the possibility of grain acceleration at young SNR shocks (throughout the free expansion and Sedov-Taylor phases) and the maximum energies reached by the accelerated grains. We investigate the potential implications on the abundance of refractory species relative to volatile elements in the cosmic-ray composition.}
   {We rely on semi-analytical descriptions of particle acceleration at strong shocks, and on self-similar solutions for the dynamics of SNR shock waves. For simplicity, type Ia thermonuclear SNRs expanding in uniform interstellar medium are considered.}
   {We find that the acceleration of dust grains at relativistic speed is possible, up to Lorentz factor of $\sim 10^{2}$, kinetic energy $E_{\rm k}/\text{nuc}\sim 10^2$ GeV/nuc for the smaller grains of size $a\sim 5 \times 10^{-7}$ cm. We find that the subsequent sputtering of grains can produce nuclei with a rigidity sufficient to be injected in the process of diffusive shock acceleration. Such scenario can help naturally account for the overabundance  of refractory elements  in the Galactic cosmic-ray composition, provided that a fraction $\eta \sim 10^{-3}-10^{-2}$ of dust grains swept up by a SNR are energized through DSA. 
   }
   {}


   \maketitle
%

\section{Introduction}
Dust grains have been shown to be potentially accelerated in interstellar shock waves through diffusive shock acceleration (DSA). 
Several works have discussed this possibility~\citep{epstein1980,ellison1997}, and how the rapid grain destruction following the acceleration could play a role in the production of cosmic rays~\citep{cesarsky1980,bibring1981}. 

In the former work of~\citet{epstein1980}, dust grains are seen as significant ions of huge mass-to-charge ratio, i.e., large rigidity. The possible acceleration is then counterbalanced by the multiple losses suffered by the dust grains~\citep{ellison1997,hoang2015}: interactions with ambient radiation fields, gas and dust, which can lead to thermal sublimation, Coulomb explosion, grain-grain collisions, or sputtering. 

As discussed in ~\citet{hoang2015}, the acceleration of dust grains to relativistic speeds is challenging (as for instance in the case of acceleration through radiation pressure), and the numerous interactions with the interstellar medium (ISM) and radiations fields are also expected to limit their survival on distance typically of $\lesssim$ pc or $\lesssim$ kpc depending on the size and composition of grains, thus limiting the possibility of detection of relativistic dust grains in our atmosphere. 

On the other hand, the acceleration of dust grains to relativistic/sub-relativistic speed is of the special interest in the context of the composition of Galactic cosmic rays. Indeed, several measurements have indicated that the refractory elements (i.e., likely to be found in dust phases, such as Mg, Al, Si, Ca, Fe, Co or Ni) are overabundant compared to volatile elements (i.e., likely to be found in gas phases). The acceleration of dust grains has been shown to potential play a role in explaining this measurement~\citep{meyer1997, ellison1997}. 
It was more specifically suggested that refractories could be injected in the ISM as  Galactic CRs from a population of energized dust grains formed in core collapse supernovae sputtered in the reverse and forward shocks~\citep{lingenfelter1998,lingenfelter2019}. However, the abundance of the refractories in the Galactic cosmic rays is somewhat similar to the one of the standard cosmic composition, which argues against the sole implication of core collapse supernovae. In particular since a substantial fraction $\sim 70\%$ of Fe originates from thermonuclear supernovae, these objects should also play a role \citep{tatischeff2021}. 

In this paper, we follow the approach of~\citet{ellison1997} to discuss the maximum momentum/velocity/kinetic energy reached by dust grains accelerated through DSA throughout the lifetime of a typical benchmark supernova remnant (SNR). We consider realistic composition and size distribution \citep{mathis1977} for dust grains in the interstellar medium and reassess the various interaction processes of the grains in SNR shocks. We then study the acceleration of refractories from sputtered energized dust grains to discuss the observed overabundance of refractory elements in the Galactic cosmic-ray composition.  

\section{Dust grains at SNR shocks}

\subsection{Physical properties}

As in~\citet{ellison1997}, we consider dust grains of characteristic size $a$ and electric potential $\phi$, so that the charge on a spherical grain is of order $q \sim 4 \pi \epsilon_0 a \phi$ ($\epsilon_0$ is the vacuum permittivity).  
This corresponds to a grain charge of: 
\begin{equation}
Q_{\rm G}= \frac{q}{e} \approx 700 \left( \frac{a}{10^{-5}~{\rm cm}} \right)  \left( \frac{\phi}{10~{\rm V}} \right)~.
\end{equation}
The number of atoms in the (spherical) grain is: 
\begin{equation}
N_{\rm G} \sim  a^3 n_d =  10^8  \left( \frac{a}{10^{-5}~{\rm cm}}\right)^3 \left( \frac{n_d}{10^{23}~\text{cm}^{-3}}\right)~,
\end{equation}
with $n_d=\rho_d/(\mu~{\rm amu})$ the atomic number density of the dust grain, $\mu$ being the average atomic weight of the grain atoms and ${\rm amu}$ the atomic mass unit (${\rm amu}=1.66\times 10^{-24}$~g). In the case of a non-spherical~\citep{min2007} or asperous grain, $N_{\rm G}$ is expected to be smaller. 
The mass of a spherical grain is: 
\begin{equation}
m_{\rm G} \sim  a^3 \rho_d = 3 \times 10^{-15} \left( \frac{a}{10^{-5}~{\rm cm}}\right)^3 \left( \frac{\rho_d}{3~\text{g cm}^{-3}}\right)~{\rm g}~.
\end{equation}
The atomic weight is $A_{\rm G}= \mu N_{\rm G}$ with $\mu$ the average atomic weight of the grain atoms. The corresponding mass--to--charge ratio is: 
\begin{equation}
\frac{A_{\rm G}}{Q_{\rm G}} \approx 1.7 \times 10^6 \left(\frac{\mu}{12} \right) \left( \frac{a}{10^{-5}~{\rm cm}}\right)^2 \left( \frac{\phi}{10~{\rm V}} \right)^{-1} \left( \frac{n_d}{10^{23}~\text{cm}^{-3}}\right)~,
\end{equation}
which that can reach values $\gtrsim 10^{5-6}$ for some dust grains, several orders of magnitude above ions. 

The considered dust grains move through the plasma at a velocity $\beta_{\rm G}$ and Lorentz factor $\gamma_{\rm G}$, and thus have a rigidity $R$: 
\begin{equation}
\begin{aligned}
R= &\frac{pc}{Q_{\rm G} e} = \frac{m_p c^2}{e} \left( \frac{A_{\rm G}}{Q_{\rm G}}\right) \beta_{\rm G} \gamma_{\rm G}  \\
&\approx 1.6 \times 10^6  \beta_{\rm G} \gamma_{\rm G}  \left(\frac{\mu}{12} \right) \left( \frac{a}{10^{-5}~{\rm cm}}\right)^2 \left( \frac{\phi}{10~{\rm V}} \right)^{-1} \left( \frac{n_d}{10^{23}~\text{cm}^{-3}}\right)~\text{GV}~,
\label{eq:rigidity}
\end{aligned}
\end{equation}
where $p$ is the grain momentum, $c$ the speed of light and $m_p$ the proton mass.

We consider a size distribution for dust grains following~\citet{mathis1977}. The number of dust grains whose size is between $a$ and $a+\text{d}a$ reads:
\begin{equation}
\text{d}n (a) \propto (a/a_0)^{-3.5} \text{d}a~,
\label{eq:mathis}
\end{equation}
normalized to that $\sim 1\%$ of the ISM mass is found in dust of sizes between $a_{\rm min}=5 \times 10^{-7}$ cm and $a_{\rm max}=5 \times  10^{-5}$ cm. 

It is usual to distinguish at least two types of dust grains in the ISM: carbonaceous and silicate grains~\citep{draine2003}, with typical properties as reported in Tab.~\ref{tab:grains}. Polycyclic Aromatic Hydrocarbons (PAHs) are not considered in the present study.

\begin{table}
  \caption{Physical properties of graphite and silicate dust grains. $n_{d}$ is the atomic number density, $\rho_d$ the mass density, and $\mu$ the average atomic weight.}
\begin{center}
\begin{tabular}{ c c c  } 
  & graphite & silicate \\ 
  \hline
$ n_{\rm d}$ [cm$^{-3}$] &  $10^{23}$ &   $10^{23}$ \\ 
 $ \rho_{\rm d}$ [g cm$^{-3}$] & 2.2 & 3.5  \\ 
  $ \mu $ & 12 & 20 \\ 
\end{tabular}
\end{center}
\label{tab:grains}
\end{table}

\subsection{Interactions of dust grains at SNR shocks}

Accelerated dust grains can interact with the various constituents of the ISM: photons, ambient gas, and dust grains. We are especially interested in the mechanisms that can lead to the destruction of accelerated dust grains.
A few relevant mechanisms affecting the survival of dust grains include: thermal sublimation, Coulomb explosion, and sputtering/grain-grain collisions. For the acceleration at SNR shock waves, sputtering and grain-grain collisions are most relevant. 

Indeed, in the case of thermal sublimation, the dust grains  heated by interactions with particles and various photon fields in the ISM  can lead to the complete destruction of the grains, but
the typical timescale associated to thermal sublimation make the process important for grain destruction if a powerful bolometric source such as a massive star is located within a few AU~\citep{guhathakurta1989,waxman2000,hoang2015}. We thus neglect this effect in this paper. 

Coulomb explosions and field emission can occur as the result of an increased charge of the  dust grains~\citep{draine1979}. The increase of the potential of the dust grains $\phi$, and tensile strength $S=(\phi/a)^2/4 \pi$ is only bearable up to a maximum  value where the grain explodes. When a dust grain  is sufficiently charged, the emission of individual ions (ion field emission) can also gradually destroy the grain without explosion.
At least two mechanisms can account for the increase of the  dust grain charge: collisional charging and photoelectric emission. Collisional charging happens in the high-energy regime when incident electrons and ions cannot be stopped or recombine in the grain but rather lead to the production of secondary electrons, which can potentially escape from the grain, thus increasing the grain charge. The photoelectric emission occurs when the absorption of high energy photons leads to the emission of primary photoelectric electrons, Auger electrons, or secondary electrons (excited by primary photoelectric electrons or Auger electrons).  Photoelectric emission is generally relevant for dust grains close to stars \citep[typically few AU;][]{hoang2015}, but it can be neglected for the accelerations of grains at SNRs. 


\subsubsection{Sputtering}
\label{sec:losses}
Sputtering  can lead to the destruction of dust grains. 
Sputtering corresponds to the ejection of matter (atoms from the dust grains)  in the energetic collision of a dust grain and a particle of the ISM. The direct \textit{knock-on} sputtering of accelerated dust grains, is efficient in the low-energy regime (E< 0.1 MeV/u). In the high--energy regime, the $\textit{electronic}$ sputtering, where the electronic excitations lead to the ejection of atoms, becomes dominant~\citep{dasgupta1979,bringa2004}. 

\citet{ellison1997}, following the approach of~\citep{dwek1987}, evaluate the loss time accounting for frictional interactions of dust grain with the background plasma (neglecting grain-grain collision), as: 
\begin{eqnarray}
\tau_{\rm loss} & \approx & \frac{A_{\rm G}}{1.4 n_{\rm H} a^2 \beta_{\rm G}c} 
\nonumber \\
& \approx & 9.2 
\left( \frac{\mu}{12} \right) \left( \frac{a}{10^{-5}~\text{cm}} \right)  \left( \frac{n_d}{10^{23}~\text{cm}^{-3}} \right)\left( \frac{n_{\rm H}}{1~ \text{cm}^{-3}}\right)^{-1} \left(\frac{\beta_{\rm G}}{1} \right)^{-1}~\text{yr}~, 
\label{eq:tau_loss}
\end{eqnarray}
where $n_{\rm H}$ is the H number density in the ambient medium.
The timescale for sputtering can be estimated by considering that about $\sim 0.5-1$ \% of collisions with ambient gas atoms leads to the sputtering of an atom from the grain surface, and that, on average, a collision reduces $A_{\rm G}$ by $\mu$: 
\begin{eqnarray}
\tau_{\rm sp} & \approx & \frac{100 A_{\rm G}}{n_{\rm H} a^2 \beta_{\rm G}c \mu}
\nonumber \\
& \approx & 10^2  
\left( \frac{a}{10^{-5}~\text{cm}} \right) \left( \frac{n_d}{10^{23}~ \text{cm}^{-3}} \right) \left( \frac{n_{\rm H}}{1~\text{cm}^{-3}}\right)^{-1} \left(\frac{\beta_{\rm G}}{1} \right)^{-1}~\text{yr}~.
\label{eq:sputtering}
\end{eqnarray}
In addition to the \textit{knock-on} sputtering, thermal sputtering can also be considered through a typical timescale~\citep{draine2011}: 
\begin{equation}
\tau_{\rm sp, thermal}\approx 10^{5} \left[  1 + \left( \frac{T}{10^6~ \text{K}}\right)^{-3} \right] \left( \frac{a}{10^{-5}~\text{cm}}\right) \left(\frac{n_{\rm H}}{1~\text{cm}^{-3}} \right)^{-1} \text{yr}~,
\end{equation}
where $T$ is the temperature of the considered plasma. This effect is significantly more important downstream in the shock-heated plasma. 

\subsubsection{Grain-grain collisions}

The timescale associated with grain-grain collisions is estimated as: 
\begin{equation}
\tau_{\rm g-g}^{-1} \approx \int_{a_0}^{a_{\rm max}} \frac{ \dd n(a)}{\dd a} \dd a \;  (a^2 +a_{\rm T}^2) \beta c~,
\end{equation}
where $a_{\rm T}$ is the typical radius of the target dust grain, the interactions with the grains of various sizes $a$ are accounted for with the integral over $a$. 
Assuming a distribution as in~\citet{mathis1977} (Eq.~\eqref{eq:mathis}), $\dd n/\dd a = n_{\rm gas} A_{\rm MRN} a^{-3.5}$ with typically 
$A_{\rm MRN} \approx 3 \times 10^{-25}$ cm$^{2.5}$ valid between $a_{\rm min}=5 \times 10^{-7} $ cm and $a_{\rm max} = 5 \times 10^{-5}$ cm, so that $\tau_{\rm g-g}^{-1} \approx \pi \beta c n_{\rm gas} (a_{\rm min}^{-0.5}/2 + a_{\rm T}^2 a_{\rm min}^{-2.5}/2.5 ) $, i.e. $\tau_{\rm g-g}$ is governed by the maximum of $\sim a_{\rm min}^{-0.5}$ and $\sim a_{\rm T}^{2} a_{\rm min}^{-2.5}$ while $a_{\rm max}$ is of little importance. 
Moreover, let us mention that other distributions of dust grains at SNR shocks (e.g., log-parabola function~\citep{bocchio2016}, or more elaborate~\citep{zubko2004,weingartner2001}) have been shown to provide good descriptions of the dust around SNRs. In the context of this paper, such distributions can potentially affect the timescales discussed above, for instance in the grain-grain collision, where the prescriptions used in Bocchio tend to increase the characteristic timescale of grain-grain interactions, but do not, however, significantly impact our results.


%


\section{Diffusive shock acceleration}

The general theory of DSA predicts that the acceleration of charged particles (and thus of dust grains in our approach) follows a power law in momentum. The distribution function of dust grains of size $a$ will read:
\begin{equation}
f (a,p,t) = A (a,t) \left(\frac{p}{p_0 }\right)^{-s} \exp \left[- \frac{p}{{p_{\rm max}(a,t)}}\right]~,
\end{equation}
where the slope in the test-particle limit is $s=3r/(r-1)=4$ for a strong shock of compression factor $r=4$ . Here, $p_0= m_{\rm p} c$, and the normalization of the spectrum of accelerated dust grains can be estimated assuming that a fraction $\xi$ of the number of dust grains of size $a$ enter DSA. In principle, this fraction can change for the grains of different sizes.
The question of how ions are injected in the DSA process is complex and still open. For protons, whose distribution in momentum is expected to follow a Maxwellian distribution, it is usual to assume that only protons with sufficient momentum are injected in DSA, with a condition that typically reads $p_{\rm inj} \approx \chi p_{\rm th}$ with $\chi \sim 3-5$~\citep{blasi2005}, and $p_{\rm th}=\sqrt{2 m k_{\rm B} T_{\rm down}}$ is the momentum at the  thermal peak of the Maxwellian downstream. As DSA is a rigidity dependent mechanism, we can assume that the same condition on the minimum rigidity is required for other ions (and dust grains) to get injected in the acceleration process. 

Still, we remark that all dust grains, in the rest frame of the plasma downstream, arrive from upstream with a velocity $v=(1-1/r) u_{\rm sh}$, where $u_{\rm sh}$ is the velocity of the SNR shock wave in the observer rest frame. Their corresponding momentum (and rigidity) with respect to the plasma downstream is $m(a)v$, largely satisfying the usual conditions on the minimum rigidity required to enter DSA cycles. We can then assume that $\eta$, the fraction of dust grains from upstream energized through DSA is the same for all dust grains.
Moreover, their temperature is expected to remain very low compared to the plasma temperature, and thus, unlike protons, the 'thermal distribution' is irrelevant for dust grains.
The number of accelerated dust grains $\text{d} N (a, t)$ of size between $a$  and $a + \text{d}a$ at a time $t$ reads: 
\begin{equation}
\text{d}N (a,t)= \int \text{d}p 4 \pi p^2 f(a, p,t) = \eta \text{d}n(a)~.
\label{eq:number}
\end{equation}

The maximum momentum of dust grains accelerated through DSA can be limited by losses suffered by the dust grains, and by their escape upstream from the acceleration region.
Following~\citet{ellison1997}, the maximum momentum reached due to losses can be estimated by equating the acceleration time $\tau_{\rm acc}$ to the minimum of the typical losses time scales and the age of the accelerator: 
\begin{equation}
\tau_{\rm acc} \lesssim \text{min}(\tau_{\rm sp}, \tau_{g-g}, t_{\rm age})~.
\label{eq:loss}
\end{equation}
The typical loss times $\tau_{\rm sp}, \tau_{g-g}$ are discussed in Section~\ref{sec:losses}.
The typical acceleration time is: 
\begin{equation}
\tau_{\rm acc} = \frac{3}{u_{\rm 1}-u_2} \int_{p^a_{\rm inj}}^p \left( \frac{D_1}{u_{1}}  +\frac{D_2}{u_{2}}\right) \frac{\text{d} p'}{p'}
\end{equation}
where $D_1, u_1$ ($D_2, u_2$) are the diffusion coefficient and velocity upstream (downstream). 

In order to investigate  the maximum energy of accelerated particles, we assume a Bohm-like diffusion coefficient
\begin{equation}
D_1= 1/3 r_{\rm L} v
\end{equation}
 where  
\begin{equation}
v = \left(1+ (m_{\rm G}c/p)^2 \right)^{-1/2} c
\end{equation}
is the velocity and
\begin{equation}
r_{\rm L}= \frac{p c}{Q_{\rm G} e B} 
\end{equation} 
the  Larmor radius of dust grains of mass $m_{\rm G}$, momentum $p$, and charge $Q_{\rm G}$, in a magnetic field $B$.

At young SNR shocks, magnetic field values of the order of few $\sim 100$ $\mu$G have been inferred from observations~\citep{vink2012}. Although the exact processes at play are not known, a consensual picture is that the streaming of accelerated particles upstream the shock excite instabilities in the plasma, thus leading to magnetic field amplification. The dominant mechanism at fast expanding shocks (young SNRs) corresponds to the growth of \textit{non-resonant} instabilities \citep[Bell modes;][]{bell2004,bell2013,schure2014}, and the corresponding magnetic field,  attained when the growth of instabilities saturate, typically reads: 
\begin{equation}
\delta B_1 = \sqrt{\frac{4 \pi \xi_{\rm CR} \rho u_{\rm sh}^2}{\Lambda} \frac{u_{\rm sh}}{c}}~,
\end{equation}
where $\rho$ is the hydrogen density in the ISM and $\Lambda= {\rm ln} (p_{\rm max}/mc)$.
The associated maximum momentum for protons is~\citep{bell2013}:
\begin{equation}
p_{\rm max} \approx \frac{r_{\rm sh}}{10} \frac{\xi_{\rm CR}e \sqrt{4 \pi \rho}}{\Lambda} \left( \frac{u_{\rm sh}}{c}\right)^2~.
\end{equation}

The maximum energy reached by dust grains can also be limited by their escape upstream from the acceleration region. In the DSA process, accelerated particles are reflected back toward the shock by scattering on turbulence generated by nonthermal particles in the precursor, mostly protons. The maximum energy of the confined grains is then determined by the lowest frequency of the excited turbulence, which corresponds to the gyrofrequency of protons at their maximum (relativistic) energy: $\nu_{\rm p}^{\rm min} =c / (2 \pi r_{\rm L})=ceB/(2 \pi E^{\rm max}_{\rm p})$. The grain confinement condition can be expressed in terms of the grain gyrofrequency as $\nu_{\rm G} > \nu_{\rm p}^{\rm min}$, which can be written in terms of Lorentz factors as $\gamma_{\rm G}<\gamma^{\rm max}_{\rm p} (Q_{\rm G} / A_{\rm G})$ or also $R_{\rm G}/\beta_{\rm G} < E^{\rm max}_{\rm p}/e$. We then get from Equation~\ref{eq:rigidity}, e.g. for silicate grains:
\begin{equation}
\gamma_{\rm G} \left( \frac{a}{10^{-5}~{\rm cm}}\right)^2 <  0.11 \left( \frac{\phi}{10~{\rm V}} \right) 
\left( \frac{E^{\rm max}_{\rm p}}{300~\text{TeV}}\right)~.
\label{eq:condition_confinement}
\end{equation}
This equation shows that relative large silicate grains with $a\gsim30-50$~nm are never confined in the shock precursor and thus cannot be accelerated by the DSA mechanism.

In addition, the acceleration can only take place provided that the gyration time of the accelerated dust grains is shorter than the acceleration time: $\tau_{\rm G} \ll t_{\rm acc}$ with:
\begin{equation}
\tau_{\rm G} = \frac{1}{\nu_{\rm G}} = \frac{2 \pi r_{\rm L}}{\beta_{\rm G}c}.
\end{equation}
\begin{figure}
\includegraphics[width=0.5\textwidth]{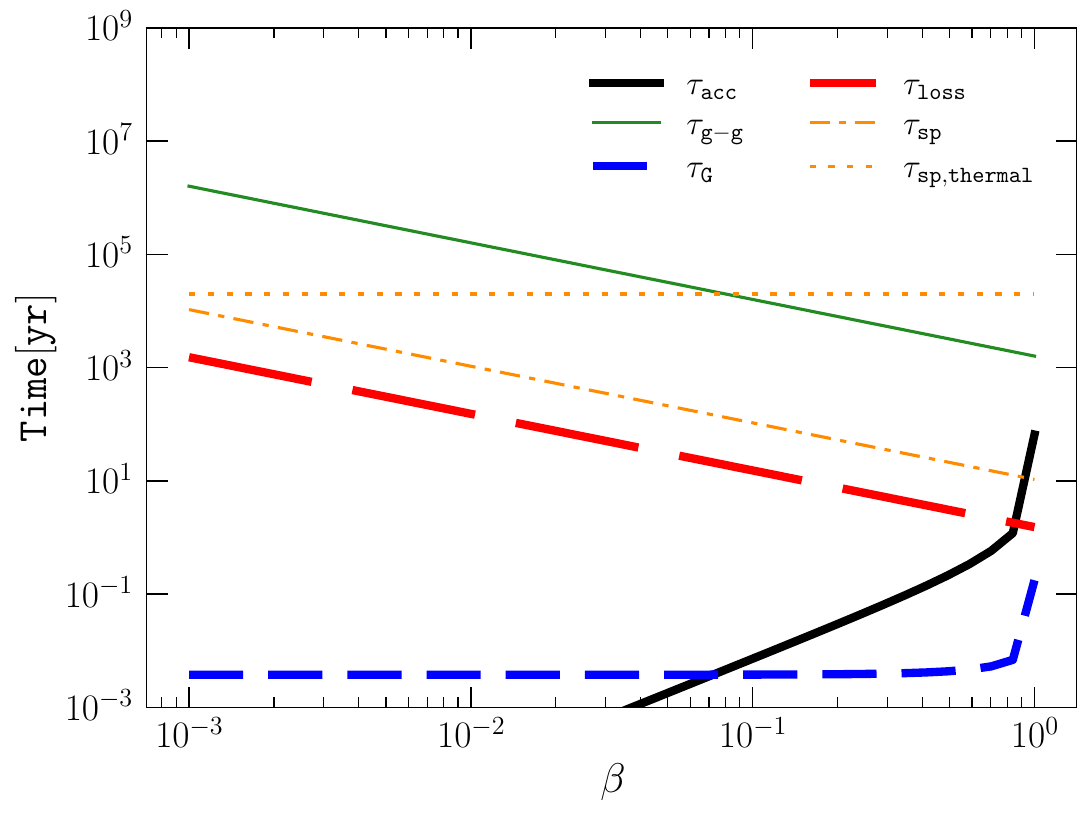}
\caption{Timescales relevant to the acceleration of a silicate dust grain of size $a=10^{-6}$ cm and shock velocity $u_{\rm sh}= 10^{9}$ cm/s vs the grain velocity $\beta$ in units of c. The relevant typical times shown are : the acceleration time $\tau_{\rm acc}$ (black solid), the grain-grain interaction time $\tau_{\rm g-g}$ (thin green solid), the gyration time (blue dashed), the grain loss time from frictional interactions with the background plasma (red dashed), the associated sputtering time (thin orange dot-dashed), and the thermal sputtering time (thin orange dotted).}
\label{fig:timescales}
\end{figure}

\begin{figure}
\includegraphics[width=.5\textwidth]{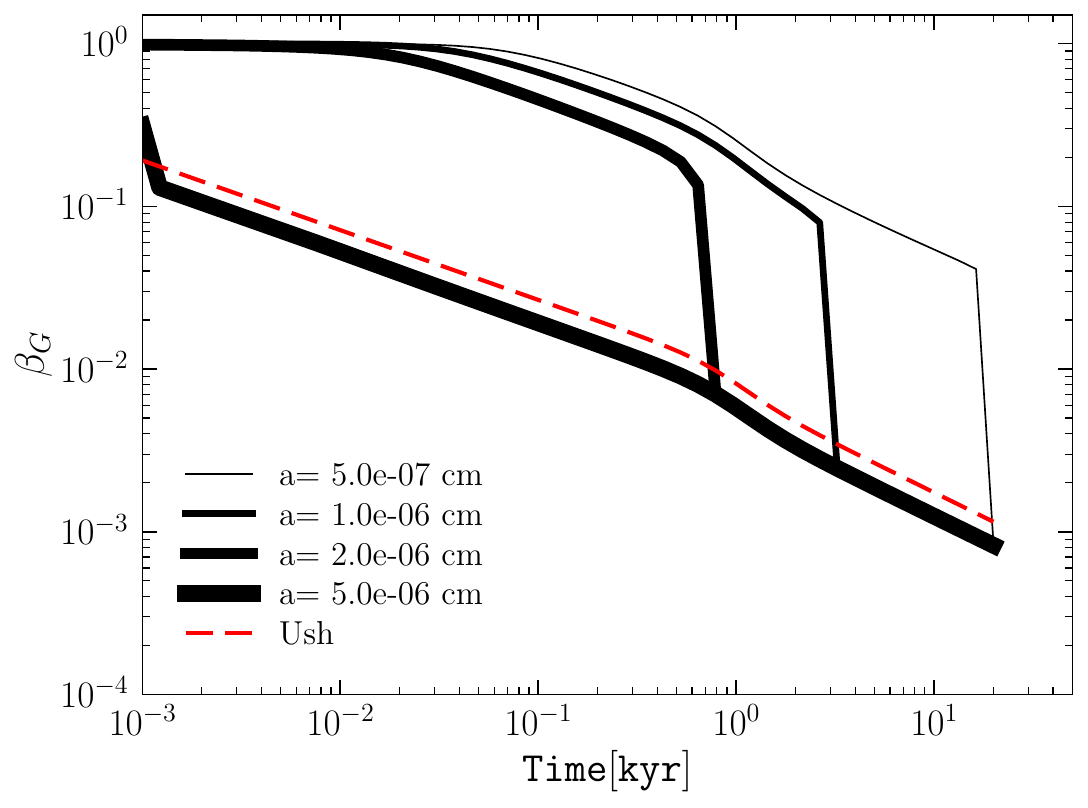}
\includegraphics[width=.5\textwidth]{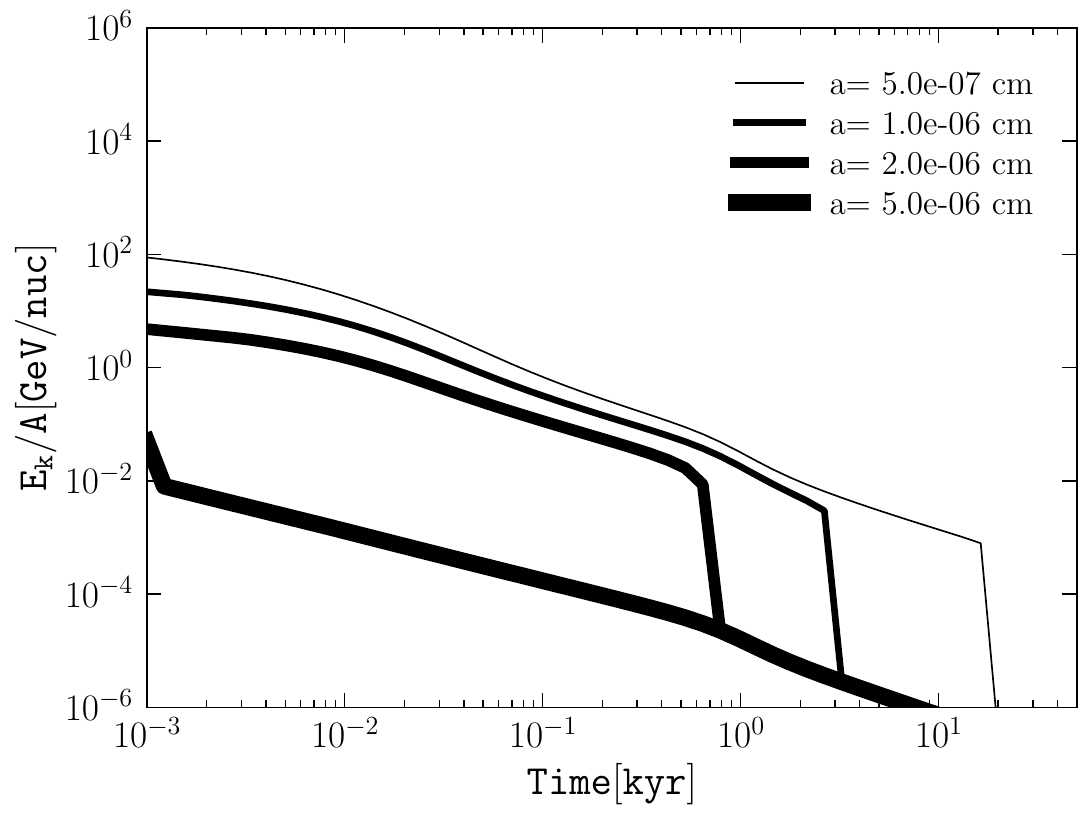}
\includegraphics[width=.5\textwidth]{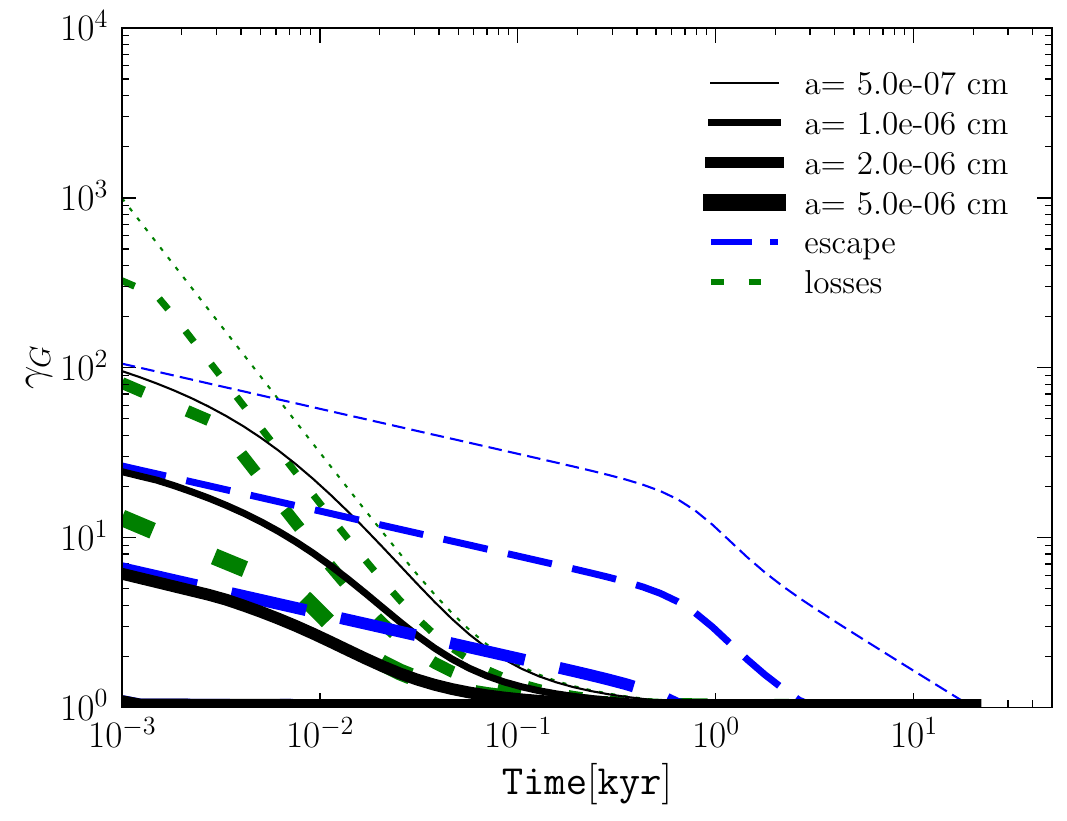}
\caption{Maximum velocity (top), kinetic energy (middle), and Lorentz factor (bottom) of accelerated silicate grains, the grain size increases from thin to thick. The red dashed curve in the top panel shows the evolution of the shock velocity. On the bottom panel, the blue dashed and green dotted lines correspond to the maximum Lorentz factor obtained with the escape condition (Eq.~\eqref{eq:condition_confinement}) and with the losses condition (Eq.~\eqref{eq:loss}), respectivly. }
\label{fig:beta_max}
\end{figure}

The timescales relevant for $E_{\rm max}$ are illustrated in Fig.~\ref{fig:timescales}, and the obtained maximum velocities, kinetic energy per nucleon and Lorentz factor are shown in Fig.~\ref{fig:beta_max} in the case of silicate grains. The maximum Lorentz factor set by the escape or by the losses considered are also plotted in Fig.~\ref{fig:beta_max} (bottom panel), illustrating that the escape is typically the most stringent condition in the first few $10^{-2}$ kyr.

Considering silicates or graphite grains does not substantially affect the quantities shown in Fig.~\ref{fig:beta_max}. However, considering a different average magnetic field value impacts the maximum energy, as for instance an average magnetic field of $\sim 3$ $\mu$G leads to a maximum Lorentz factor of $\sim 3$, and kinetic energy of $\sim 5$ GeV/nuc for the smaller grains of $a=5 \times 10^{-7}$ cm. 

\begin{figure}
\includegraphics[width=0.45\textwidth]{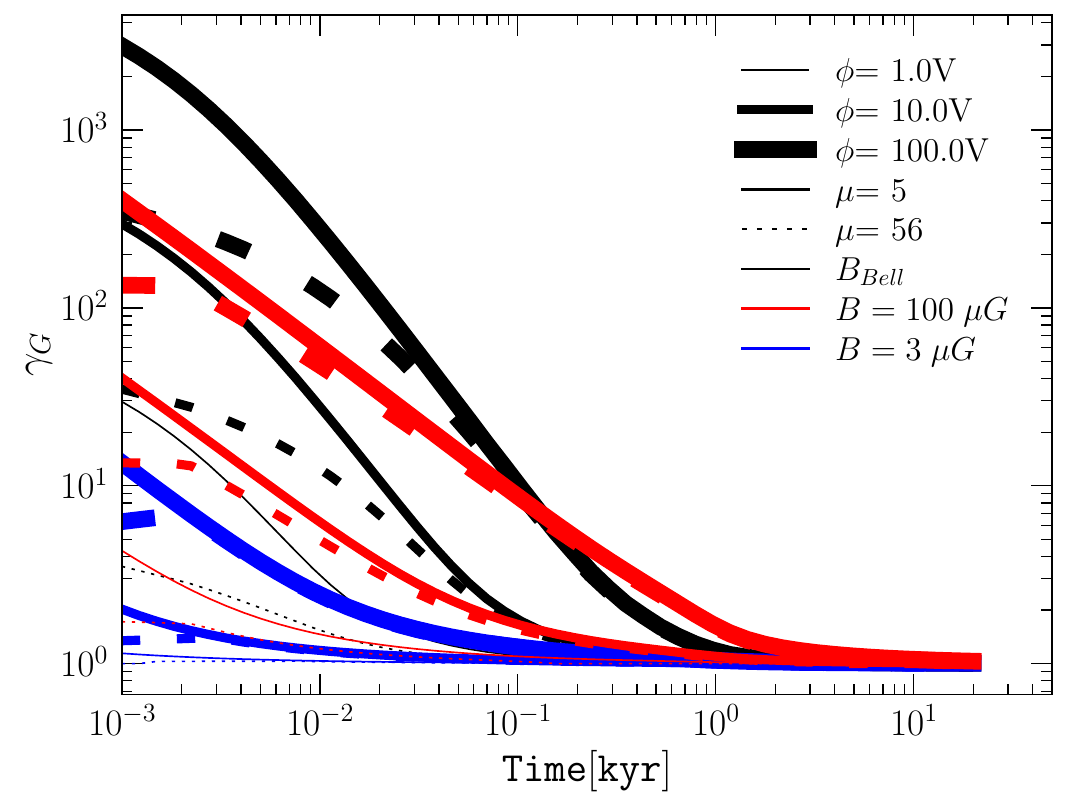}
\caption{Maximum Lorentz factor of dust grains with $a=5 \times 10^{-7}$ cm for various grain average atomic weight and potential. The thickness of the lines accounts for the change in potential (from 1~V to 100~V). The black, red and blue lines correspond respectively to a magnetic field amplified due to the growth of non-resonant streaming instabilities (Bell), to a fixed value of 100 $\mu$G and a fixed value of 3 $\mu$G. The solid lines assume a mean mass $\mu =5$, and the dashed lines $\mu =56$.}
\label{fig:emax_phi}
\end{figure}

The influence of the average magnetic field, as well as the grain average atomic weight $\mu$ and the potential $\phi$ on the maximum Lorentz factor of the accelerated dust grains is additionally illustrated in Fig.~\ref{fig:emax_phi}.

\begin{figure}[h]
\includegraphics[width=0.49\textwidth]{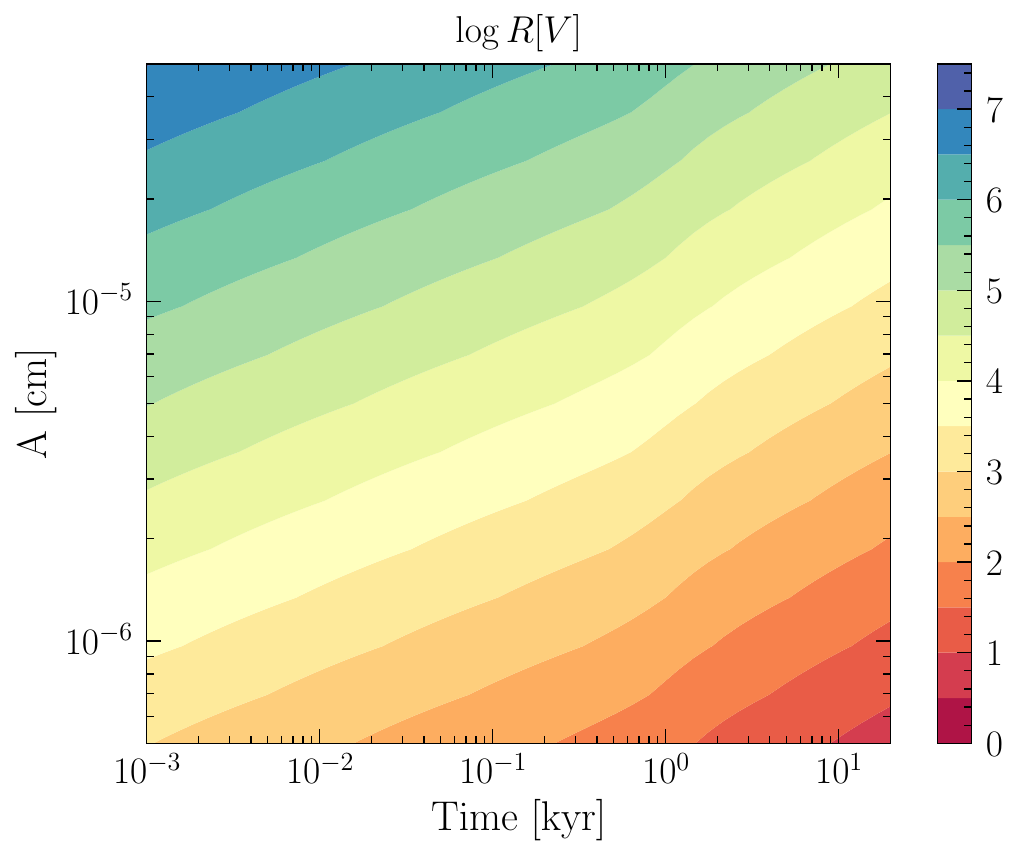}
\includegraphics[width=0.49\textwidth]{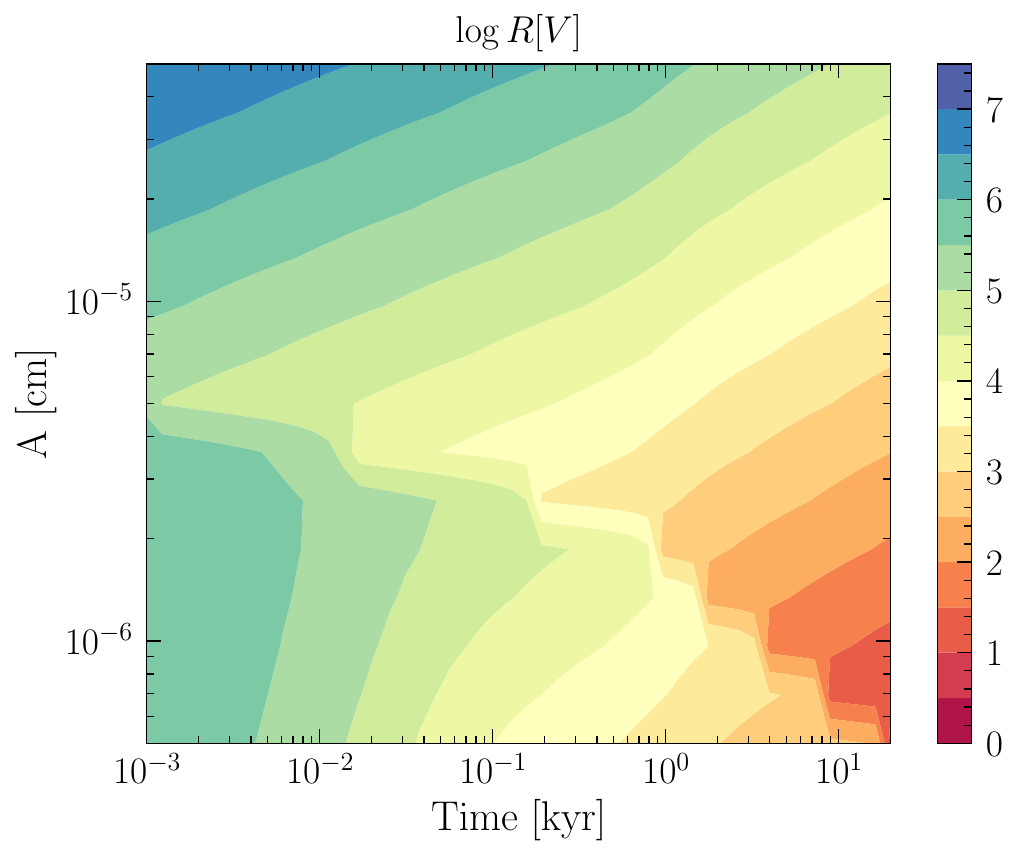}
\caption{Rigidity of silicate dust grains arriving downstream (top panel) and at maximum energized through DSA (bottom panel).}
\label{fig:contour_rigidity_grains_1}
\end{figure}

As mentioned previously, in the rest frame of the downstream plasma, the grains, if assumed to be advected in the plasma, arrive with a velocity $(1-1/r) u_{\rm sh}$,  and thus with large momenta (rigidities) compared to ion in the plasma. The rigidity of the dust grain arriving from the upstream plasma, and the maximum rigidity of the dust grain that underwent DSA are shown in Fig.~\ref{fig:contour_rigidity_grains_1} in the case of silicate grains.

\section{The abundance of refractory elements}
A substantial fraction of dust grains in a SNR environment is expected to be destroyed~\citep{slavin2015}. This is also expected for accelerated dust grains, for which sputtering is expected to play an important role. Interestingly, the atoms sputtered from the accelerated dust grains have a higher rigidity than the atom sputtered from the thermal grains, or the grains simply advected in the plasma. 
We  examine the case of silicate dust grains, under the hypothesis that all Si atoms in the ISM are locked in such grains. The  Si abundance in CRs compared to that in the solar system (SS) composition has been found to be typically $\sim 20$ times greater than the  CR-to-SS abundance ratio for protons~\citep{tatischeff2021}. The acceleration and sputtering of dust grains in SNR shocks has been proposed as a possible way to explain the observed overabundance of refractory elements in the CR composition. 

Focusing on the case of silicates, we compute throughout the lifetime of a SNR (up until the end of the Sedov Taylor phase) the amount of accelerated protons throught DSA, under the usual assumption that at each time a fraction of the ram pressure is converted into CR protons $P_{\rm CR, p}= \xi_{\rm p} \rho u_{\rm sh}^2(t)$ with $\xi_{\rm CR}\sim 0.1$, that corresponds to a cumulative ratio of accelerated-to-swept-up protons of $\eta_{\rm p} \sim 10^{-5}$ at the end of the Sedov Taylor phase.

\begin{figure}[h]
\includegraphics[width=0.49\textwidth]{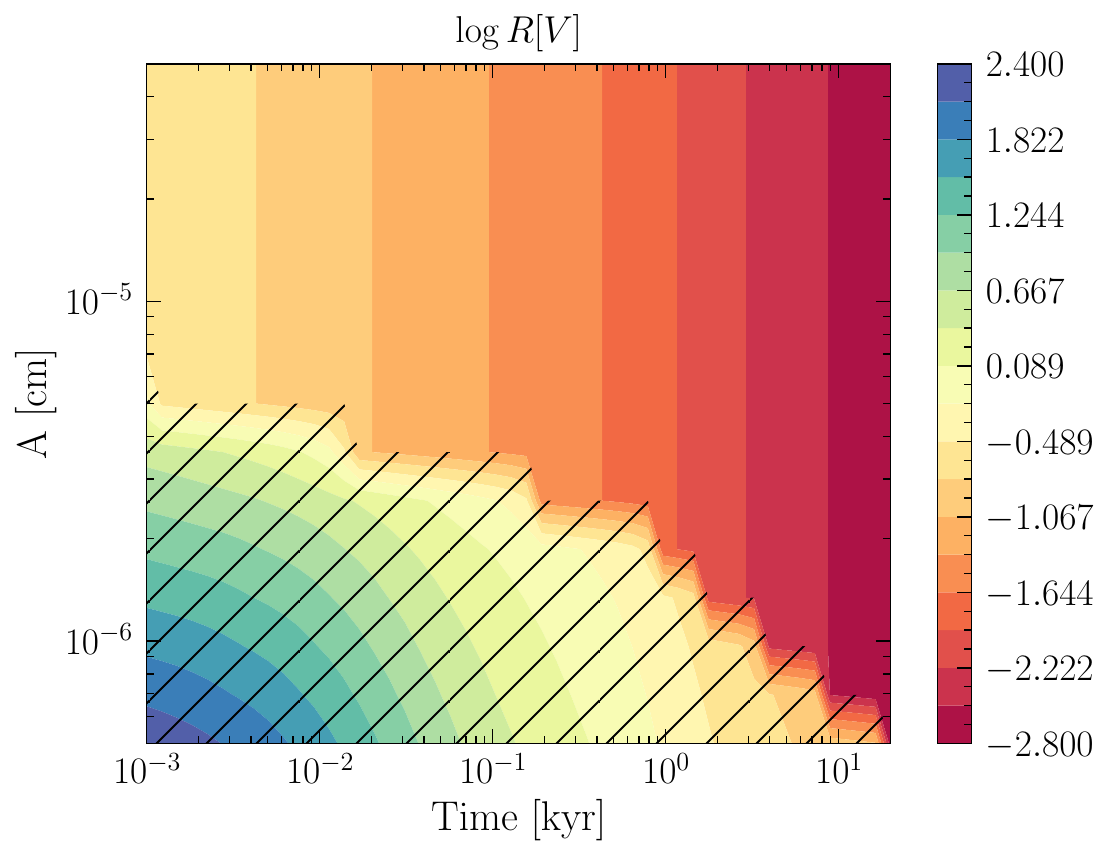}
\caption{Rigidity of silicon sputtered from accelerated silicate dust grains. The black hatched area corresponds to the region where the rigidity of the sputtered silicon is lower than the rigidity required for the injection into DSA.}
\label{fig:contour_rigidity_grains}
\end{figure}

Additionally, we compute the amount of dust grains accelerated, and number of sputtered silicon with rigidity larger than the minimum injection ridigity of the protons required to enter DSA: these silicon nuclei are thus considered to have a sufficient rigidity to be accelerated through DSA and contribute to the CR composition. The typical rigidity of the sputtered silicon is illustrated in Fig.~\ref{fig:contour_rigidity_grains}, showing that throughout the entire SNR lifetime, all dust grains typically smaller than $\sim 4 \times 10^{-6}$ can be sputtered to produce silicon that can get into DSA.  
At a given time $t$, the amount of sputtered silicon is estimated by comparing the sputtering timescale given in Eq.~\eqref{eq:sputtering} to the dynamical timescale at a given time $t_{\rm dyn}=\min (t, t_{\rm adv})$, where $t_{\rm adv}\sim D_1/u_{\rm sh}^2$ with $D_1$ the diffusion coefficient of dust grains upstream. 
In such approach, we implicitly consider that all the silicon sputtered upstream will be advected to the shock, and enter DSA provided that the rigidity satisfies the criterion on injection. The sputtering need to occur upstream of the shock so that the sputtered ions can be swept into the shock for subsequent acceleration by DSA. The silicon sputtered downstream is quickly advected downstream and lost inside the SNR. Ionization losses for silicon are also taken into account through their typical they timescale as in~\citet{ellison1997} but do not significantly affect our calculation.

\begin{figure}[h]
\includegraphics[width=0.49\textwidth]{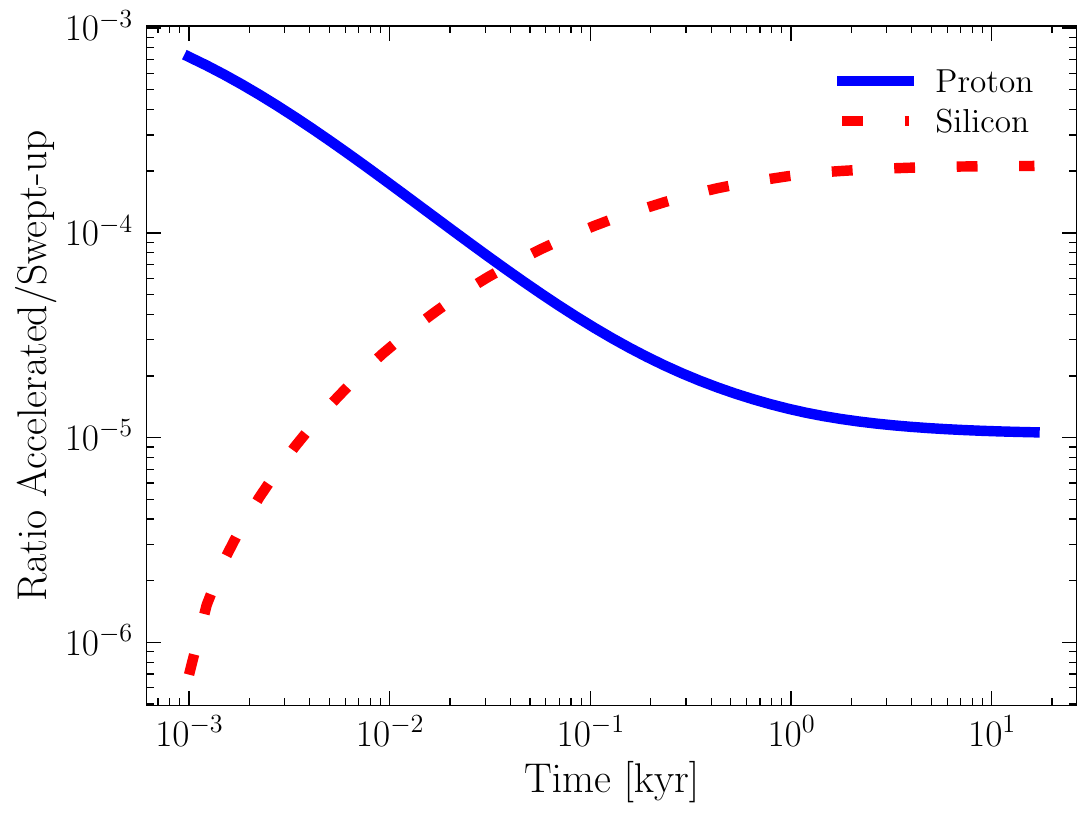}
\caption{Ratio of the cumulative accelerated ions to cumulative swept-up ions. The blue solid line corresponds to protons,  the red dotted to silicon considering sputtering as main destruction mechanism for the dust grains. The calculation carried with the number of CRs/dust grains ($\eta_{\rm CR}= 10^{-5}$, and $\eta_a \sim 5\times 10^{-3}$ in order to get a ratio silicon/protons of $\sim 20$) at the end of the Sedov-Taylor phase. }
\label{fig:ratio}
\end{figure}

We now estimate the cumulative ratio of accelerated-to-swept-up silicon and protons  denoted $R_{\rm si}$ and $R_{\rm p}$. The main free parameter in our calculation is the fraction of dust grains energized through DSA $\eta$. The scaling of $R_{\rm si}$ is proportional to $\eta$, and $R_{\rm p}\propto \eta_{\rm CR}$, so that in order to obtain a ratio $R_{\rm si}$/$R_{\rm p} \sim 20$ assuming a typical value of $\eta_{\rm CR} \sim 10^{-5}$, the efficiency of acceleration of dust grains needs to be of the order of $\eta \sim 5 \times 10^{-3}$. In other words: 
$R_{\rm si}$/$R_{\rm p} \sim 20 \left( \frac{\eta}{5 \times 10^{-3}} \right) \left( \frac{\eta_{\rm CR}}{10^{-5}} \right)^{-1}$ 
(as shown in Fig.~\ref{fig:ratio}). 

Under the assumption that the increase fraction of silicon in CRs comes from accelerated and sputtered dust grain, an efficiency of acceleration of dust grains $\eta \sim 10^{-3}-10^{-2}$ is required, i.e. 2 to 3 orders of magnitude more than for protons for which $\eta \sim 10^{-5}$. The need for an increased efficiency for dust grains is easily understood since as discussed in the previous Section, only smaller dust grains can be accelerated through DSA ($\beta_{\rm max}$ larger than the shock velocity), and thus a large amount of accelerated small dust grains is needed, so that these dust grains can subsequently be sputtered.

Such increased efficiency is easily accounted for given that the dust grain arriving from upstream have rigidities vastly larger than the minimum rigidity required for protons to enter DSA. 
In this calculation, the typical total ratio of the accelerated dust silicates to the silicates swept-up by the SNR shock is found of the order of $\sim 2 \times 10^{-4}$, that corresponds to an efficiency of $\eta_a \sim 5 \times 10^{-3}$ where the grains smaller than $\lesssim 5 \times 10^{-6}$ cm are efficiently energized through DSA. This amounts to an acceleration of $\sim 5 \times 10^{-3}$ of the total mass of dust grains swept-up by the SNR throughout its active life. 

The results of our calculation is illustrated in Fig.~\ref{fig:ratio}. In our simplified approach, all CR Si ions originate from the sputtering of accelerated silicate grains, which are assumed to be all destroyed at the end of the adiabatic phase ($\sim$ 20 kyr).
All the non-accelerated silicates are also substantially destroyed from various processes, as often at least $\sim$ 50\% are estimated to be destroyed at SNR shocks from various processes~\citep{zhu2019}, but the rigidity of silicon arriving downstream. Non-accelerated silicates are also substantially destroyed at SNR shocks from various processes \citep{zhu2019}, but the rigidity of Si ions sputtered from non-accelerated silicates in the downstream medium is smaller than the rigidity required to enter into the DSA process. 

Let us mention that for this calculation, we have normalized the number of accelerated grains through $\eta$ (Eq.~\eqref{eq:number}), but it is also possible to normalize $f(a,p,t)$ in pressure,  assuming that a fraction $\xi_{a}$ of the \textit{partial} pressure for grains $n(a) m(a) \dd a u_{\rm sh}^2$ is accelerated through DSA. The normalization in pressure easily allows to obtain $R_{\rm si}$/$R_{\rm p} \sim 20$ provided $\xi_a \sim 4 \times 10^{-3} \left( \frac{\xi_{\rm CR}}{0.1} \right)$. 

However, several aspects could strongly affect the efficiency of acceleration of dust grains. The charging of dust grains -- for instance through collisional charging~\citep{mckee1987,weingartner2006} -- can also play a significant role on the efficiency of acceleration, and an increase of a factor $\sim~100$ in the charge leads a grain of size $a = 5 \times 10^{-7}$~cm to a rigidity  smaller than the one required for injection into DSA. Second, as discussed in Section~\ref{sec:losses}, the sputtering rates are not well constrained, and any increase in the sputtering timescale would allow for larger values of $\eta$.

\section{Conclusions}

DSA at  SNR shocks is found to be able to energize dust grains up to  relativistic energies, e.g., $\gamma \sim 10^{2}$, $E_{\rm k} \sim 10^{2}$ GeV/nuc for grains of size  $a \sim 5 \times 10^{-7}$ cm. The maximum energy of small grains with $a \lesssim 5 \times 10^{-6}$ cm is limited by the losses/destruction at play. For grains larger than $a \gtrsim 5 \times 10^{-6}$, the escape of the grains from the shock precursor prevents their acceleration.

Additionally, the  sputtering of ions from accelerated dust grains can help account for the overabundance of refractories in CRs, as the enhanced rigidity of sputtered ions allow them to be in turn injected in DSA more efficiently. A silicon-to-proton abundance ratio of $\sim 20$, as found in the measurement of CRs \citep{tatischeff2021} can be accounted for, provided that the efficiency of acceleration of dust grains is typically of the order of $\eta_{a}\sim 10^{-3}-10^{-2}$, i.e. 2-3 orders of magnitude larger than the one for protons $\eta_{\rm CR} \sim 10^{-5}$. 

Several effects require further investigations to clearly understand the details of the acceleration of dust grains, such as, 1) the importance of dust grain destruction in the SNR environment, and especially close to the shock discontinuity (i.e. sputering, destruction by charging,...); 2) the injection of dust grains in the DSA process; 3) the charging of the dust grains that can affect both the injection and the maximum rigidity; 4) non-linear effects of DSA on the acceleration of grains, and the potential consequences of grain acceleration on the acceleration of other ions present in the plasma; 5) the importance of the grain distribution in size, e.g., the presence of large dust grains ($\sim 10^{-4}$ cm) was shown to potentially be reflected and accelerated in the preshock gas before being destroyed, thus possibily affecting the shock environment in which CR/dust grain acceleration is taking place~\citep{slavin2004}; 6) the case of remnants from core-collapse supernovae, expanding in very diverse environments, needs also to be taken into account.

\begin{acknowledgements}
PC acknowledges support from the GALAPAGOS PSL Starting Grant. The authors warmly thank J. Raymond and J. Slavin for fruitful discussions and constructive comments on the manuscript. 
\end{acknowledgements}

%
   \bibliographystyle{aa} 
   \bibliography{dust} 
%

\end{document}